# Deterministic generation and switching of dissipative Kerr soliton in a thermally controlled micro-resonator

Zhizhou Lu,[1,2,†] Weiqiang Wang,[1,2,*,†] Wenfu Zhang,[1,2,*] Sai T. Chu,[3] Brent E. Little,[1] Mulong Liu,[1,2] Leiran Wang,[1,2] Chang-Ling Zou,[4,5] Chun-Hua Dong,[4,5] Bailing Zhao,[1,2] And Wei Zhao[1,2]

[1]State Key Laboratory of Transient Optics and Photonics, Xi'an Institute of Optics and Precision Mechanics (XIOPM), Chinese Academy of Sciences (CAS), Xi'an 710119, China
[2]University of Chinese Academy of Sciences, Beijing 100049, China
[3]Department of Physics and Materials Science, City University of Hong Kong, Hong Kong, China
[4]Key Laboratory of Quantum Information, CAS, University of Science and Technology of China, Hefei, Anhui 230026, China
[5]Synergetic Innovation Center of Quantum Information & Quantum Physics, University of Science and Technology of China, Hefei, Anhui 230026, China
*Corresponding author: wwq@opt.ac.cn, wfuzhang@opt.ac.cn

**Dissipative Kerr solitons (DKSs) in the high-Q micro-resonator correspond to self-organized short pulse in time domain via double balance between dispersion and Kerr nonlinearity, as well as cavity loss and parametric gain. In this paper, we first experimentally demonstrate deterministic generation and switching of DKSs in a thermally controlled micro-ring resonator based on high-index doped silica glass platform. In our scheme, an auxiliary laser is introduced to timely balance the intra-cavity heat fluctuation. By decreasing the operating temperature through a thermo-electric cooler, primary-, chaotic- comb and soliton crystal are firstly generated, then increasing the temperature, DKSs switching and single soliton are robustly accessed, which is independent of the tuning speed. During the switching process, varieties of DKSs are identified by tens of the featured discrete "soliton-steps", which is favorable for study of on-chip soliton interactions and nonlinear applications.**

http://dx.doi.org/10.1364/OL.99.099999

Micro-resonator-based dissipative Kerr solitons (DKSs) via parametric four-wave mixing (FWM) have drawn considerable interests in the last decade for broad bandwidth, low power consumption and miniaturization operation [1, 2]. DKSs have been successfully generated in $MgF_2$, $Si_3N_4$, Si, silica and AlN platforms [3-7], among which the spectra cover from visible to mid-infrared window and the repetition rate ranges from several gigahertz (GHz) to terahertz (THz) [8]. Up to now, DKSs have revolutionized the fields of dual-comb spectroscopy [9, 10], chip-scale ranging system [11, 12], low-noise microwave generation [13], optical coherent communication system [14, 15], etc.

To realize practical applications, stably and deterministically accessing single-soliton state is the fundamental issue. The generation of DKSs in a high-Q micro-cavity requires keeping the pump at effective red-detuned regime where the system suffers from thermal instability [16]. Therefore, complicated techniques are developed to enable DKSs formation, such as rapid frequency tuning or two-step "power kicking" protocol [3, 6], as well as rapid thermal-tuning methods [17]. However, the number of soliton formed in the resonator is difficult to control until the backward frequency scheme is introduced by H. Guo, *et. al* [18]. Regrettably, these schemes rely on strict timing sequence in which the tuning mode and speed have to be carefully designed, leaving it difficult to implement. Recently, an auxiliary-laser-heating method is proposed to access and stabilize the DKS state [19, 20]. Within this scheme, an auxiliary laser is introduced to balance the thermal fluctuation inside the resonator. The intra-cavity heat reduction accompanying with the soliton state switching is compensated by the auxiliary laser which locates at the blue-detuned regime. Thus, the thermal instability is effectively alleviated and arbitrary laser detuning is allowed for the pump, which is favorable for the research of soliton behavior of MRR comprehensively. However, this method still relies on wavelength tunable laser which suffers from relatively high noise and broader linewidth typically on the order of 100kHz. Furthermore, the auxiliary and pump laser are in the same polarization state, thus the subsequent spectra-separation remains challenging.

In this paper, we demonstrate deterministic DKSs generation and switching in a high-index doped silica glass micro-ring resonator (MRR) [21-23] using thermal tuning method, which has been proven to realize robust soliton crystal (SC) as presented in our previous work [24], however, transition to multi- and single soliton is prohibited using single pump in our follow-up attempts. Here we introduce an auxiliary laser which is counter-coupled into the four-port MRR and orthogonally polarized with the pump. By sequentially decreasing and increasing the operating temperature, DKS switching is observed. Furthermore, it is convenient to separate the DKS from reflected auxiliary comb components using a polarization beam splitter (PBS), which could enable DKS-based applications.

Figure 1 shows the experimental set up for DKSs generation and switching behavior, as well as spectra-separation. The monolithic integrated four-port high-Q MRR is packaged in a butterfly-package with a thermo-electric cooler (TEC), as shown in the inset, where a Chinese 1 dime is for scale. The ring radius is ~592.1µm, corresponding to the free spectral range (FSR) of ~49 GHz [24, 25]. The cross-sections of the ring and bus waveguides are both 2µm × 3µm. The MRR exhibits anomalous dispersion in communication band [26]. The Q factors of $TM_{00}$ mode (pump) and $TE_{00}$ mode (auxiliary) are $2.05 \times 10^6$ and $1.69 \times 10^6$, respectively [26]. The operating temperature of the MRR can be precisely tuned through an external TEC controller. The pump laser ($TM_{00}$, ECDL #1) for soliton is a narrow-linewidth laser whose wavelength is 1561.792nm in our experiment while the auxiliary laser ($TE_{00}$, ECDL #2) is wavelength-tunable, allowing at least 1×FSR tuning range (0.4nm for this device). The two lasers are boosted to a similar power level (34.5~36.6dBm in our experiment) using two commercial high-power erbium doped fiber amplifiers (EDFAs). Two fiber polarization controllers (FPCs) are used to control the polarization states of the two lasers. The circulators inhibit the strong light from transmitting to the EDFAs. The generated DKSs coupled out from the drop port are measured using two optical spectrum analyzers (OSAs), electrical spectrum analyzer (ESA) and oscilloscope (OSC).

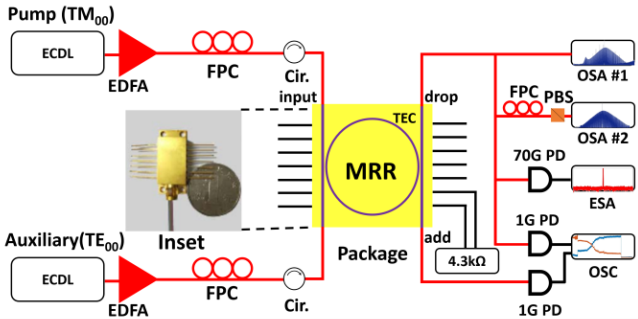

**Fig. 1.** Experimental set up for DKSs generation, switching behavior and combs-separation. The inset is the image of the packaged MRR used in the experiment. EDFA: erbium-doped fiber amplifier. FPC: fiber polarization controller. Cir: high power circulator. MRR: micro-ring resonator. PBS: polarization beam splitter. OSA: optical spectrum analyzer. OSC: oscilloscope. PD: photodetector. ESA: electric spectrum analyzer.

In our experiments, the auxiliary and the pump laser are coupled into the orthogonal modes of the MRR simultaneously through carefully tuning the wavelength of the auxiliary laser, such as 1558.26nm in one of our tests. The pump laser is slightly closer to the corresponding resonance compared to the auxiliary laser while both of them are blue-detuned relative to each resonance. Fig. 2(a) sketches the complete evolution process of the intra-cavity optical spectra. Decreasing the operating temperature through the TEC, the pump first reaches the threshold of optical parametric oscillation (OPO). The primary and modulation instability (MI) combs are successively generated while the auxiliary laser keeps approaching the resonance but under the OPO threshold value (state (i) and (ii) of Fig. 2(a)). The corresponding experimental optical and electrical spectra are presented in state (i) and (ii) of Fig. 2(b), respectively. Further decreasing the temperature, the resonance climbs over the pump and the red-detuned regime is achieved, the corresponding optical spectrum exhibits SC feature (state (iii) of Figs 2(a) and (b)) [24, 27]. During this process, the resulting power drop of pump comb inside the MRR is effectively compensated by the power increment of auxiliary laser which locates at blue-detuned regime. This is clearly proved by the complementary steps of power traces shown in Figs. 3(a) and (b), which are monitored via the drop- and add-port of the MRR. Further decreasing the temperature (forward tuning), the frequency combs will vanish as illustrated in Fig. 3(a). In contrast, increasing the temperature from this stage (backward tuning), the soliton switching could be stably accessed. This behavior is indicated in Fig. 3(b) where the number of solitons in the MRR is reduced step by step until single soliton state is achieved. State (iv) of Fig. 2 (b) shows the single soliton spectrum, where the $sech^2$ fitting (red) suggest 4.4nm (11×FSR) Raman self-frequency shift (RSFS) [28] and the corresponding RF spectra show ~48.97 GHz of repetition rate. Meanwhile, the discrete spikes comb lines reveals that the primary comb is stimulated simultaneously by the auxiliary laser. It should be noted that the state of the auxiliary comb is determined by the relative position of the pump and auxiliary lasers. The two combs are easily separated using a PBS in subsequent experiments.

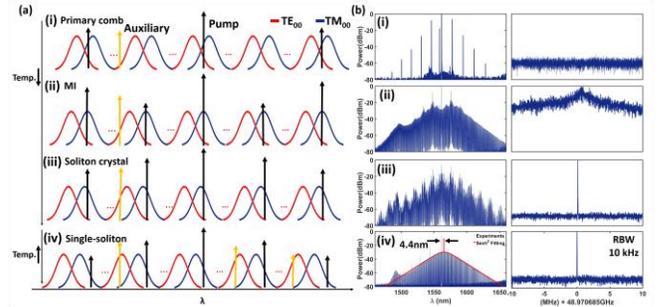

**Fig. 2.** DKS evolution with optical spectra and RF beat note. (a) Schematic diagram of DKS evolution. Decreasing the temperature (Temp.↓) from (i) primary comb, (ii) chaotic MI comb to (iii) SC state, then increasing the temperature ((Temp.↑)) until (iv) single soliton state is achieved. The corresponding optical and electrical spectra are shown in (b). The repetition rate of the single soliton is ~48.97 GHz with 4.4nm RSFS. The pump and auxiliary laser are fixed while the resonances are shifted during the thermal tuning process.

It is intriguing that when the operating temperature is forward and backward tuned, the intra-cavity optical field acts entirely different. Such distinct feature is attributed to thermal nonlinearity in the MRR. Specifically, to maintain the soliton states, the frequency of the pump should keep in the soliton existence range (SER) ($\delta_L < \delta < \delta_H$), where $\delta_L$ and $\delta_H$ are the lower and upper boundaries of the SER, respectively. The upper boundary $\delta_H$ is degenerate with respect to the number of solitons [18]. Therefore, DKSs switching can't be realized by increasing the effective detuning of the pump, which corresponds to decreasing the operating temperature. On the contrary, the lower boundary of the SER is nondegenerate with respect to the number of solitons which benefits from the thermal nonlinearity [18] of the high-Q MRR. A staircase patterned power trace can be expected when the pump sweeps backward, equivalent to increasing the temperature in our experiments [18]. Each step corresponds to a specific soliton state, as an example, the markers in Fig.3 (b) are three- and four- soliton states whose spectra are shown as (i) and (ii) in Fig. 3(c), respectively.

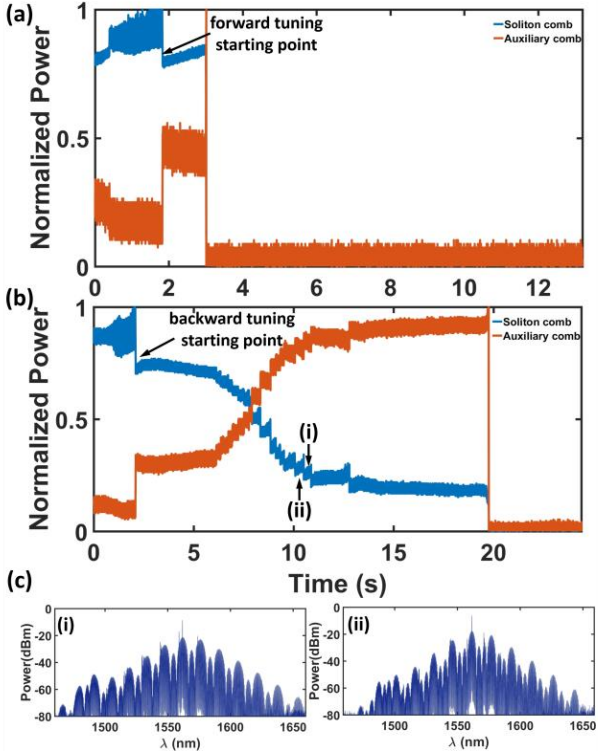

**Fig. 3.** The power trace under forward (a) and forward-backward tuning (b) schemes, the discrete steps suggest different soliton states, such as three- (i) and four- (ii) soliton state, whose spectra are shown as (i) and (ii) of (c), respectively.

In the optical spectrum of single soliton shown in Fig. 2(b), the spatial-mode-interaction induced dispersive waves (DWs) [29] are clearly observed at the wavelength around 1490nm. We characterize the DW using a PBS, the reflected auxiliary and the pump comb spectra are presented in Figs. 4(a) and (b). It is clearly seen that the components of the DWs are almost in the same polarization with the auxiliary comb (TE mode). To qualitatively investigate this situation, further dispersion calculations are performed [29] and shown in Fig. 4(c), where unperturbed (dashed line, no mode crossing) and perturbed (solid line, mode crossing) dispersion curves are both presented for comparison. The calculation and experiment reveal that the $TE_{01}$ mode is involved as interacting mode and strong mode-interaction [30] between $TE_{01}$ and $TM_{00}$ mode occurred inside the MRR, which leads to an avoided mode crossing on the mode dispersion curve (dashed line) that induces the DWs and mode hybridizations [24]. Therefore the DW in our device is expected to be in $TE_{01}$ mode. The DW wavelength deviation between calculation and experiment is attributed to be thermo-optic effect.

To determine the Raman property as well as characterizing the temporal profile of the single soliton, we performed the simulation using the damped, driven Lugiato-Lefever equation (LLE) [2, 31-32]:

$$(t_R \frac{\partial}{\partial t} + \frac{\alpha+\kappa}{2} + i\frac{\beta_2 L}{2!}\frac{\partial^2}{\partial \tau^2} + i\frac{\beta_3 L}{3!}\frac{\partial^3}{\partial \tau^3} + i\delta)E(t,\tau)$$
$$-\sqrt{\kappa}E_{in} = i\gamma L(|E(t,\tau)|^2 + \tau_R \frac{\partial |E(t,\tau)|^2}{\partial \tau}) \quad (1)$$

where $E(t,\tau)$ is the complex slowly varying amplitude of the intra-cavity electric field, $t$ and $\tau$ are slow and fast time of the system. $\alpha$ and $\kappa$ corresponds to the loss and coupling parameter, $\beta_2$ is the GVD value and $\beta_3$ is third order dispersion. $L$ is the total MRR length. $\delta$ and $\gamma$ are detuning and nonlinear parameters, respectively. First term on right hand side is Kerr nonlinearity while another contributes to the Raman interactions, where $\tau_R$ is the Raman shock time which is partly responsible for the amount of self-frequency shift [31, 32]. During the simulation, $\beta_3$ and $\delta$ related with each comb line around DW position are used to adjust the global dispersion curve.

During the simulation, $\tau_R$ is 2.7fs [31] and other material-related parameters are calculated from a finite element solver and are listed in the caption. Taking the RSFS and wavelength-dependent loss into Fourier terms [31] and applying fourth-order Runge-Kutta method [33], the simulated spectrum (red envelope) in Fig. 4(d) is in good agreement with the experiment results where the inset shows the simulated soliton pulse profile exhibiting 114 fs of full width at half maximum (FWHM).

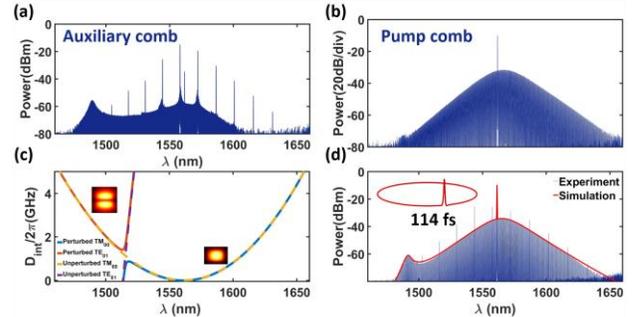

**Fig. 4.** DWs characterization and simulation. Reflected auxiliary (a) and pump (b) comb after PBS separation. DWs in same polarization with auxiliary comb. (c) Simulated integrated dispersion curves for $TM_{00}$ and $TE_{01}$ mode families, which qualitatively explains the DW formation. Dash and solid lines shows unperturbed and perturbed modes, respectively. (d) Simulated (red) and experimental (midnight blue) spectra, showing good agreement. The inset is the simulated soliton pulse, the FWHM is 114fs. The parameters in simulation: $\beta_2$ = -0.033 ps²/m, $\beta_3$ = -0.00395 ps³/m, $\alpha$ = 0.0059, $\kappa$ = 0.0015, $L$ = 0.0037 m, $\gamma$ = 0.1102/ (m×W).

By leveraging LLE, we have also investigated two-soliton state with different soliton pulse distribution along the MRR, which could be directly inferred from the specific optical spectra. Figure 5 shows the typical experimental (midnight blue) and simulated (red) spectra which suggest the angle differences between the two pulses of 173.65°, 18°, 180°, and 8.36°, respectively. In particular, Fig. 5(c) corresponds to 180° with the comb spacing of 2×FSR (~98GHz, for our device), and Fig. 5(d) with the angle difference of 8.36°, which is the smallest angle difference observed in our device. The diversity of the soliton spectra may help us to study the soliton interactions inside a micro-cavity system.

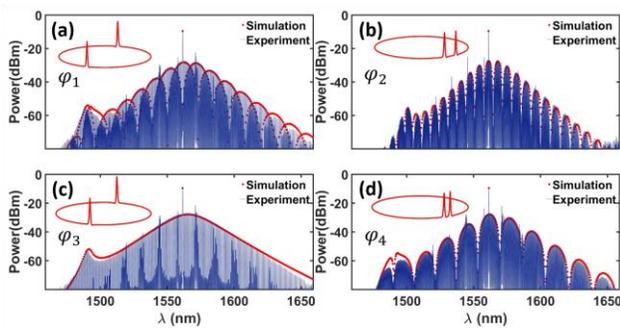

**Fig. 5.** Different spectra in two-soliton state with specific angle differences of (a) 173.65°, (b) 18°, (c) 180°, and (d) 8.36°. Red: simulation. Midnight Blue: experiments.

Realizing soliton switching and deterministic single-soliton micro-comb in a micro-resonator is of great importance for exploring the applications and intra-cavity dynamics of DKSs. With the help of the auxiliary laser, the pump in arbitrary SER can be stably achieved and different states of DKS can be generated. Right now, the number of soliton is not strictly reduced one by one in our experiments (N, N-1⋯1) [18]. This is because that manually temperature-tuning through a variable-resistor in our current configuration is not precise. We believe that deterministically one-by-one soliton annihilation can be realized once the temperature tuning precision is further improved. This can be realized using programmable digital-to-analog converter to adjust the TEC temperature instead of inexactly manual tuning, besides, an on-chip micro-heater could also be a considerable choice for deterministic DKSs switching.

In conclusion, we experimentally demonstrate deterministic generation and switching of DKSs in a thermally controlled high-index doped silica MRR with wavelength-fixed pump. Our scheme is absolutely tuning-speed independent which attributes to the fact that the intra-cavity heat fluctuation is timely balanced by an orthogonally polarized auxiliary laser. Varieties of DKSs are obtained, which is insightful to study on-chip soliton interaction. Furthermore, the auxiliary laser could be completely filtered out simply using a PBS. Our scheme is expected to provide high-performance DKSs in practical applications.

**Funding.** This work was supported by the National Natural Science Foundation of China (NSFC) (Grant No. 61635013, 61675231, 61475188, 61705257), the Strategic Priority Research Program of the Chinese Academy of Sciences (Grant No. XDB24030600). C.-L. Zou and C.-H. Dong acknowledge the National Key Research and Development Program of China (Grant No. 2016YFA0301303).

†These authors contributed equally to this work.